# IRONMAP: Iron Network Mapping and Analysis Protocol for Detecting Over-Time Brain Iron Abnormalities in Neurological Disease

Jack A. Reeves[1], Fahad Salman[1], Michael G. Dwyer[1,2], Niels Bergsland[1], Sarah Muldoon[3], Bianca Weinstock-Guttman[4], Robert Zivadinov[1,2], Ferdinand Schweser[1,2]

[1]Buffalo Neuroimaging Analysis Center, Department of Neurology, Jacobs School of Medicine and Biomedical Sciences, University at Buffalo, State University of New York, Buffalo, NY, USA; [2]Center for Biomedical Imaging at the Clinical Translational Science Institute, University at Buffalo, State University of New York, Buffalo, NY, USA; [3]Department of Mathematics, University at Buffalo, State University of New York, Buffalo, NY, USA; [4]Jacobs Neurological Institute, Buffalo, NY, USA.

## Abstract

Pathologically altered iron levels, detected using iron-sensitive MRI techniques such as quantitative susceptibility mapping (QSM), are observed in neurological disorders such as multiple sclerosis (MS) and may play a crucial role in disease pathophysiology. However, brain iron changes occur slowly, even in neurological diseases, and can be influenced by physiological factors such as diet. Therefore, novel analysis methods are needed to improve sensitivity to disease-related iron changes as compared to conventional region-based analysis methods. This study introduces IRONMAP, Iron Network Mapping and Analysis Protocol, which is a novel network-based analysis method to evaluate over-time changes in magnetic susceptibility. With this novel methodology, we analyzed short-term (<1 year) longitudinal QSM data from a cohort of individuals with MS (pwMS) and healthy controls (HCs) and assessed disease-related network patterns, comparing the new approach to a conventional per-region rate-of-change method. IRONMAP analysis was able to detect over-time, MS-related brain iron abnormalities that were undetectable using the rate-of-change approach. IRONMAP was applicable on the per-subject level, improving binary classification of pwMS vs HCs compared to rate-of-change data alone (areas under the curve: 0.773 vs 0.636, p = 0.024). Further analysis revealed that the observed IRONMAP-derived HC network structure closely aligned with simulated networks based on healthy aging-related susceptibility data, suggesting that disruptions in normal aging-related iron changes may contribute to the network differences seen in pwMS. IRONMAP is generalizable to any neurological disease, including Alzheimer's disease and Parkinson's disease, and may allow for study of brain iron abnormalities over shorter timeframes than previously possible.

# Introduction

Maintaining brain iron homeostasis is critical for healthy brain function.[1] Adequate iron levels are necessary for normal metabolic processes, including myelination and neurotransmitter synthesis, while excessive (improperly sequestered) iron can lead to the formation of reactive oxygen species with harmful effects.[2–4] Several factors having been linked to altered iron homeostasis in the brain, including healthy aging,[5] clinical factors such as body mass index,[6] and neurological disorders like Alzheimer's disease and multiple sclerosis (MS).[7,8] Various cellular and biochemical mechanisms have been proposed to contribute to these iron alterations, such as altered iron transport across the blood-brain barrier, inflammatory activity, or iron depletion from the glial syncytium leading to decreased concentration.[8,9]

Iron levels in the brain can be assessed non-invasively using iron-sensitive MRI techniques like quantitative susceptibility mapping (QSM), which quantifies the magnetic susceptibility of tissues,[10–12] a metric for the magnetizability. In healthy individuals, MRI measurements have shown that aging-related deep gray matter (DGM) iron dynamics vary between regions. For example, iron levels in the globus pallidus rapidly increases in adolescence and plateau around age 30, whereas iron accumulates in the caudate at a slower pace throughout the human lifespan.[13,14] These MRI findings are consistent with post-mortem histological iron measurements.[5] These findings suggest that certain DGM regions, such as the caudate and hippocampus which both accumulate iron slowly over time,[14] may have similar iron dynamics and may even share common iron transport mechanisms. This latter idea is supported by recent study showing that brain iron can be directly translocated between brain regions via axons.[15]

In people with MS (pwMS), MRI studies have shown increased iron concentrations in DGM regions such as the putamen and caudate, and decreased concentrations in the pulvinar of the thalamus.[16–18] DGM iron alterations detected on MRI predict clinical disability, disease subtype, and disease duration, independent of atrophy and white matter (WM) lesion load.[19] Due to a potential role of pathologically increased DGM iron in MS disease progression, iron chelation is current under investigation as treatment for MS progression. A limitation to observing over-time DGM iron alterations is that DGM iron levels slowly and can be influenced by confounding physiological factors such as diet.[20,21] Therefore, monitoring longitudinal iron changes requires relatively large cohorts and long follow-up times, even for group-level analyses.[16,17] Development of novel analysis methods with improved sensitivity to over-time DGM iron abnormalities are therefore needed for translation of DGM iron as a clinical neuroimaging marker.

We recently reported a network-based method which leveraged independent component analysis (ICA) to identify covarying patterns ("networks") of susceptibility change in pwMS and healthy controls (HCs).[22] Our network approach improved sensitivity in detecting MS-related magnetic susceptibility alterations compared to conventional per-region susceptibility analysis.[22] Similarly, Ravanfar et al. applied a network approach to detect susceptibility alterations in people with schizophrenia.[23] Together, these studies show that network analyses may useful in detecting disease-specific iron alterations,[22,23] and may improve sensitivity as compared to the conventional method of evaluating susceptibility levels in each region separately.[22] Additionally, as shown by Wang et al.,[15] comparing iron dynamics between regions may provide valuable insight into healthy brain iron physiology and disease pathophysiology. A limitation of these previous network-based approaches is that they were only applied to cross-sectional data. Additionally, comparisons were performed between-subjects. Therefore, the utility of network-based analyses in detecting over-time iron alterations in individual subjects is unknown.

In the present work, we introduce IRONMAP, Iron Network Mapping and Analysis Protocol. IRONMAP is a novel network analysis approach for studying over-time magnetic susceptibility changes. We hypothesized that this new method exposes short-term (< 1 year) disease-specific iron alterations in pwMS undetectable with conventional longitudinal per-region methods. We tested the hypothesis by analyzing QSM data from a cohort of pwMS and comparing the resulting IRONMAP-derived network patterns to those observed in HCs. We assessed whether values obtained from the IRONMAP approach improved classification of *individual* subjects as pwMS vs HCs, as compared to per-region rates of susceptibility change alone. Finally, we explored the physiological underpinnings of the observed IRONMAP-derived HC network patterns. To do so, we used numerical simulation to test whether the in vivo HC network patterns were similar to patterns expected from normal aging-related iron changes.

## Methods

### Participants and data collection

This study included previously-collected data identified in our imaging database of IRB-approved studies. Subjects were included if they had at least three MRI scans on the same 3T MRI scanner within one year that included the same 3D gradient-echo sequences (GRE), and either had clinically definite MS (pwMS) or were neurologically normal (i.e. healthy controls; HCs). Subjects were excluded if at least one of the identified scans was deemed unusable due to excessive motion or other artifacts. Exactly three MRI scans per subject were included for analysis, with the most recent three being selected for analysis if the subject had additional scans that fit the inclusion criteria.

Written informed consent was obtained from all participants according to the Declaration of Helsinki. Demographic and clinical data were collected during an in-person interview and with additional standardized questionnaires. Information on gadolinium administrations in pwMS was collected via retrospective evaluation of electronic medical records.

## MRI Acquisition, Reconstruction, and conventional ROI analysis

The imaging protocols included both a spoiled 3D GRE and high-resolution 3D T1-weighted (T1w) imaging. GRE imaging parameters were as follows: matrix size of 512x192x64 mm, voxel size of 0.5x1x2mm³, flip angle of 12°, TE/TR of 22ms/40ms, and bandwidth of 13.89kHz. T1w imaging parameters were: FOV 256x192 mm, isotropic 1mm resolution, TE=16 ms, and TR=600 ms. Susceptibility maps were reconstructed from raw GRE k-space data using scalar phase matching,[24,25] path-based phase unwrapping,[26] background field removal by solving the Laplacian boundary value (LBV) problem and Superfast Dipole Inversion (SDI).[27,28] Susceptibility maps were then whole-brain referenced. We segmented 5 DGM regions in both hemispheres (for each individual subject) using FSL FIRST after registering the T1w images to the susceptibility maps: caudate, hippocampus, pallidum, putamen, and thalamus. The quality of segmentations was assessed by an experienced neuroimaging researcher (J.R.). Subsequently, the ROI volumes and mean susceptibility values were calculated for each timepoint for each participant. A representative susceptibility map and segmentation are shown in Fig. 1A and 1B, respectively.

## IRONMAP Methodology

The standard ROI-based approach can assess over-time changes in susceptibility in individual regions. Here, we generalize this approach toward assessing the relationship of over-time susceptibility changes *between* regions. Specifically, we described the temporal dynamics of the susceptibility as a weighted graph in which each node represents an anatomical region and the connections between the nodes, or "edges", carry a weight that is defined by strength of the over-time correlation of the susceptibility between two regions. We calculated the 45 unique edge weights corresponding to region pairs generated by Pearson-correlating the region-average susceptibility values of the two anatomical regions across timepoints, as illustrated in Fig. 1C and 1D. We visualized group-averaged weighted graphs as correlation matrices, as shown in Fig. 1F.

## IRONMAP Analysis of Aging-Related Susceptibility Networks (in silico)

We investigated if aging-related changes in susceptibility could explain the IRONMAP-derived in vivo network structure. We based these simulations on the aging trajectories of magnetic susceptibility previously published for five bilateral DGM structures (caudate, hippocampus, pallidum, putamen, and thalamus).[14] We determined for each subject and timepoint the putative magnetic susceptibility values for each region using the age of the subject at the time of the scan. From these values, we determined the expected over-time change of the regional susceptibilities from the baseline to each follow-up timepoint. These changes were added to each subject's observed baseline susceptibility to simulate short-term susceptibility values for each DGM region. We systematically investigated the effect of random noise by adding zero-mean Gaussian noise to the simulated susceptibility values with varying standard deviations from 0.01 ppb to 1.0 ppb in steps of 0.01 ppb.

We then performed IRONMAP analysis on the simulated data, generated weighted subject graphs, and compared the simulated graphs to the graphs observed in vivo. These last steps were repeated with simulated values generated using zero-mean Gaussian noise to determine the specificity of our comparisons for age-related changes, as opposed to correlated random noise. The simulated aging analysis and noise-only analysis were repeated 1000 time each with randomly varying Gaussian noise.

## Statistics

Statistical analyses were conducted using MATLAB R2019b unless otherwise stated. Subject age was compared between HC and pwMS groups using two-tailed independent-samples T-tests, and sex was compared using a chi-squared test. Statistical significance was considered at alpha < 0.05 for all analyses.

*Removing the confounding effects of volume and gadolinium accumulation*

Prior to network analysis, linear regression was used to regress out the effect of volume on mean susceptibility for each region (across all scans) and the resulting residual values were saved. Subsequently, to control for potential confounding effects of gadolinium accumulation from contrast agent injections in the patient group,[29–31] per-region regression models were fit on the pwMS residuals using the number of gadolinium administrations since baseline scan as a predictor variable. The volume-corrected susceptibility residuals for the HC group and the volume- and gadolinium-corrected susceptibility residuals for the pwMS group were used in subsequent network analyses.

*Baseline and longitudinal DGM susceptibility comparisons*

Baseline average susceptibilities were compared between pwMS and HCs using two-tailed independent-samples T-tests.

For longitudinal comparisons, the rate of susceptibility change in each region for each subject was calculated by fitting linear regression models with susceptibility residuals as outcome variables and per-visit subject ages as predictor variables. The rates of susceptibility change were the age beta coefficients (slopes) obtained from the regression. Beta coefficients were compared between pwMS and HCs using two-tailed independent-samples T-tests. Additionally, the rates of susceptibility change for each group (pwMS and HCs separately) were tested for non-zero change using two-tailed one-sample T-tests.

Note that this regression procedure was selected in favor of calculating pre-to-post changes in order to incorporate all data into the estimation of susceptibility changes, and to ensure that rate-of-change vs network analyses comparisons used similar data (see "Comparison of subject group classification improvement using network model").

*Comparison of In Vivo Susceptibility Network Dynamics Between PwMS and HCs*

Prior to comparisons, each correlation coefficient was transformed to a z-score using the Fisher z-transformation (i.e. $z = \text{arctanh}(r)$). Values of r=1 and R=-r were set to 0.99 and -0.99 prior to the z-transformation to avoid undefined values (i.e. infinity and negative infinity).

The number of numerically negative z-transformed correlation coefficients (of the 45 unique region-pairs) were compared between pwMS and HCs using chi-squared tests. The average difference in z-scores was compared between pwMS and HCs by calculating the pwMS and HC z-score averages for each unique region-pair, subtracting the HC averages from the pwMS averages, and performing a two-tailed one-sample T-test on the 45 mean-differences. This latter analysis was repeated using the absolute values of the z-scores, to investigate whether observed differences in node strengths between groups were due to differences in correlation magnitude (e.g. correlation coefficients of 0.6 vs 0.2) or differences in correlation sign (e.g. correlation coefficients of 0.2 vs -0.2).

*Comparing Classification of pwMS and HCs Using Susceptibility Rates vs the IRONMAP Approach*

Binary regressions and receiver operating characteristic (ROC) analysis was used to test whether the correlations from the network approach added additional information for classifying subjects as pwMS or HCs, as compared to only the rates of per-region susceptibility change. This analysis was performed using SPSS version 29.0 (IBM, Armonk, NY, United States).

An initial "Rate Only Model" model was fit with subject group as the outcome variable and the 10 per-region (i.e. five bilateral DGM structures) rates of susceptibility change added as forced entry predictor variables. A second "Rate + Network Model" was then fit which in which the 45 unique network region-pair correlations were added using forward selection (at $p < 0.05$), along with the 10 per-region rates of susceptibility change as forced entry predictor variables. The predicted mean response of both final models was saved and used to generate ROC curves. Paired-sample area-under-the-curve (AUC) tests were then used to compare the ROC curve generated from the "Rate + Network Model" to the "Rate Only Model". Additionally, z-transformed correlation coefficients for the network region-pairs in the final "Rate + Network Model" were compared between pwMS and HCs using two-sided independent-samples T-tests.

*Comparison of In Vivo and Simulated (In Silico) IRONMAP-Derived Networks*

We quantified the similarity between the observed HC in vivo network and the simulated (in silico) HC aging networks by correlating their group-average matrix elements. We hypothesized that regions with similar aging-related iron dynamics, such as the hippocampus and caudate, would exhibit high in silico correlations. If aging-related iron changes was closely related to the in vivo network behavior, we expected region pairs with strong in silico correlations to also show strong in vivo correlations. Conversely, if aging-related iron changes had little relation with the in vivo network, we anticipated weak or no correlation between the in vivo and in silico networks.

For the in vivo network, each unique matrix element (n = 45) was averaged across HCs. For each noise level (n = 101) and each simulated network iteration (n = 1000), each unique in silico matrix element was averaged across HCs. Pearson correlations were then calculated between the in vivo and in silico matrix elements. For each noise level, the correlation coefficients were averaged across the n = 1000 iterations. This procedure was also applied to noise-only matrices to determine if the observed network patterns could be explained by correlated Gaussian noise.

## Results

*Demographic characteristics*

The database search identified 99 pwMS (baseline age = 43.7 ± 11.2 years, 66.7% female) and 29 HCs (baseline age = 44.1 ± 15.6 years, 72.4% female) who met the inclusion criteria and were included in subsequent analyses. There were no significant differences between pwMS and HCs in baseline age (44.1 years for HC vs 43.7 years for MS, p = 0.764) or sex (p = 0.654). Details on age, sex, disease duration, and clinical disability as assessed by the Expanded Disability Status Scale (EDSS) are provided in Table 1.

Of the 297 MRIs from pwMS, 248 were from observational studies, 35 were from prospective drug trials (26 interferon beta-1a, 5 glatiramer acetate, 3 fingolimod, and 1 teriflunomide), and 14 were from a prospective neurovascular surgery study. All HC MRI scans (n = 87) were acquired as non-disease controls from MS observational studies. All scans were acquired between 2008 and 2016.

*Baseline and longitudinal DGM susceptibility comparisons*

Details on mean DGM baseline susceptibilities and longitudinal susceptibility changes for pwMS and HCs are shown in Table 2. At baseline, pwMS had higher mean susceptibility in the left caudate (48.3 ± 14.6 ppb for pwMS and 41.3 ± 16.7 ppb for HCs, $p = 0.032$), right caudate (47.9 ± 15.8 ppb for pwMS and 39.2 ± 17.1 ppb for HCs, $p = 0.011$), left pallidum (115.5 ± 29.2 ppb for pwMS, 92.9 ± 28.1 ppb for HCs, $p < 0.001$), and right pallidum (107.6 ± 30.8 ppb for 82.5 ± 29.8 ppb for HCs, $p < 0.001$). None of the DGM regions had significantly non-zero rates of susceptibility change in either the pwMS or HC group ($p > 0.1$), and all rates of were similar between pwMS and HCs ($p > 0.25$).

*Comparison of In Vivo Susceptibility Network Dynamics Between PwMS and HCs*

The group-average pwMS IRONMAP-derived matrix for in vivo susceptibility correlations is shown in Fig. 2A, the group-average HC matrix is shown in Fig. 2B, and the pwMS-minus-HC subtraction matrix is shown in Fig. 2C. Of the 45 unique region-pairs, the group-average correlation coefficients for pwMS were numerically negative in 14/45 (31.1%) pairs. In contrast, HCs had 6/45 (13.3%) numerically negative pairs, which was significantly fewer than the pwMS (chi-squared p-value = 0.043).

In the pwMS-minus-HC subtraction matrix, 35 (78.8%) of region-pairs were numerically negative. Across all unique region-pairs, average z-transformed correlation coefficient values were lower in pwMS compared to HCs (mean for pwMS = 0.08 ± 0.13, mean for HCs = 0.18 ± 0.15, $p < 0.001$). Additionally, the average z-transformed correlation coefficient magnitude was lower in pwMS compared to HCs (mean for pwMS = 0.66 ± 0.03, mean for HCs = 0.68 ± 0.05, $p = 0.008$).

*Comparing Classification of pwMS and HCs Using Susceptibility Rates vs the IRONMAP Approach*

Figure 3 shows ROC curves for pwMS vs HC classification using the Rate Only Model and for the final Rate + Network Model. The final Rate + Network Model included correlation coefficients between the left caudate and left pallidum, the left caudate and right caudate, and the right pallidum and right thalamus, in additional to the 10 per-region susceptibility rate changes. The AUC for the final Rate + Network Model (AUC = 0.773) was significantly greater than the AUC for the Rate Only Model (AUC = 0.636, p = 0.024).

Compared to HCs, pwMS had significantly smaller z-transformed correlation coefficients between the left caudate and left pallidum ($z = 0.51 \pm 057$ for HCs and $z = 0.22 \pm 0.68$ for pwMS, p = 0.017), between the left caudate and right caudate ($z = 0.52 \pm 0.48$ for HCs and $z = 0.17 \pm 0.74$ for pwMS, p < 0.001), and between the right pallidum and right hippocampus ($z = 0.38 \pm 0.59$ for HCs and $z = 0.09 \pm 0.71$ for pwMS, p = 0.004).

*Comparison of In Vivo and Simulated (In Silico) IRONMAP-Derived Networks*

Figures 3A compares the unique group-average matrix elements of the HC in vivo network with corresponding elements of the simulated aging network for different noise levels. The correlation between in vivo and simulated values peaked at a noise level of 0.07 ppb with R = 0.600. At this noise level, the simulated aging-related susceptibility changes were able to explain $R^2$ = 36.0% of the observed temporal network dynamics. In contrast, the maximum absolute correlation observed with the noise-only simulation was R = 0.015 (Figure 3C), equivalent to < 0.1% of explained variance. Scatter plots showing the relationship between the unique matrix elements of the group-averaged HC in vivo network and the corresponding elements of the simulated aging network (noise level of 0.07 ppb) are shown in Fig. 3B, and between the group-averaged HC in vivo network and noise-only simulated network in Fig. 3D.

## Discussion

In this study, we introduced IRONMAP, Iron Network Mapping and Analysis Protocol, which is a novel network analysis method for quantitative susceptibility mapping. Importantly, we found that IRONMAP analysis improved detection of disease-specific susceptibility alterations (i.e. classification of pwMS vs HCs) as compared to use of per-region rates-of-change. Comparison of our results to *in silico* numerical simulations showed substantial overlap between susceptibility network dynamics in HCs and network behavioral expected by healthy aging. Together, these results support IRONMAP as a sensitive method for study brain iron dynamics in healthy physiology and neurological disease.

At baseline, we found higher susceptibility in the caudate and pallidum in pwMS compared to HCs. However, we did not detect significant 1-year rates of susceptibility change in either the pwMS or HC group in any DGM region, nor any rate differences between the groups. These findings are in line with previous studies showing that progressive brain iron changes occur slowly.[13,16,17] Therefore, although QSM provides a highly accurate measurement of magnetic susceptibility,[32] measurement of progressive small over-time changes may be confounded by physiologic fluctuations in brain iron, i.e. slight variations due to diet and lifestyle factors.[20,21] Fluctuations may influence the DGM itself or impact other brain regions used for QSM referencing (in our case, whole-brain referencing). In contrast, IRONMAP does not rely on progressive brain iron changes, but instead on relative changes between brain regions. This negates fluctuations caused by reference variations. This feature allowed us to detect widespread susceptibility alterations in pwMS compared to HCs in the absence of any detectable over-time changes.

We found that between-region correlations generally decreased in magnitude in pwMS compared to HCs, and that there were a higher number of numerically negative correlations in pwMS. Our numerical simulations and previous *in vivo* results provide possible insight into the mechanisms underlying these findings. The group HC correlation matrix showed substantial agreement with correlation matrices generated from simulated aging data. Specifically, DGM regions that were predicted to be highly correlated due to similar aging-related brain iron dynamics, i.e. caudate and the hippocampus, also had higher observed correlations in the *in vivo* data, and regions with dissimilar aging iron dynamics had weaker *in vivo* correlations. In contrast, the noise-only simulations did not show any relationship with the in vivo data, indicating that correlated random noise did not explain the *in vivo* findings. These finding indicates that much of the observed deviation seen in the pwMS network may be due to a breakdown of normal aging-related brain iron dynamics.

In a previous mouse study, Wang et al. showed that iron levels are negatively related between certain pairs of DGM regions, i.e. iron chelation in the thalamus leads to *increased* iron levels in the substantia nigra. The authors interpreted this as certain areas providing "negative feedback" to other brain regions, possibly as a mechanism to maintain healthy brain iron homeostasis. In our data, we found an increased number of numerically negative between-region correlations in pwMS compared to HCs. Together, these results may indicate that pathological brain iron changes in different brain regions may directly interact, rather than occurring independently. If true, this may help explain why some brain regions (e.g. caudate) have increased brain iron in MS while other regions (e.g. pulvinar) have decreased iron. Future studies analyzing brain iron feedback on animal models, such as experimental autoimmune encephalitis in mice, would provide additional information on this topic.

IRONMAP classification of subjects as pwMS and HCs (i.e. at the per-subject level) as compared to conventional per-region methodology. Importantly, our approach was not specifically optimized for detecting pwMS vs HC differences. It is therefore likely that our approach could achieve greater group separation through improvements such as finer gray matter segmentation (e.g. thalamic subnuclei), increasing the number of MRI scans included in the network analysis, and including non-DGM regions, e.g. cortical regions, which are also known to have susceptibility alteration in neurological disease.[23] Further, use of QSM preprocessing methods that attenuate the confounding effects of myelin, such as chi-separation,[33] could provide increase the sensitivity of our method to specifically detect brain iron alterations. Although the susceptibility correlation methodology was applied to pwMS in the current study, it could also be applied to study other diseases with brain iron abnormalities, such as Parkinson's disease and Alzheimer's disease.[34]

*Limitations*

One limitation for interpretation of our results is that QSM signals are affected by both paramagnetic iron, which increases susceptibility relative to water, and diamagnetic myelin, which reduces susceptibility.[35,36] This is particularly relevant in our pwMS cohort, because both iron and myelin may be altered in neurodegenerative diseases. Therefore, our results need to be confirmed in future pathohistological studies, i.e. in mice. Alternatively, future human studies could incorporate QSM pre-processing techniques, such as chi-separation,[33] to estimate the effects of iron and myelin separately on susceptibility levels.

Another confounding factor is atrophy, which may lead to increased iron concentrations if the same amount of iron is located in a smaller, atrophied region.[9] Additionally, although not intrinsic to the brain, administration of gadolinium as an MRI contrast has also been shown to affect QSM signal.[29] Gadolinium is regularly administered in pwMS to evaluate the presence of acutely-appearing lesions and has been shown to accumulate in the brain tissue.[30,31] We controlled for these potential effects by regressing out regional brain volumes (on the whole cohort) and the number of gadolinium administrations over the study interval (in pwMS). However, this approach is imperfect and these factors may still have influenced our results.

Finally, the current study required subjects to have at least three MRI scans within one year. Many pwMS and HCs in our local database had MRI scans acquired at yearly (or less frequent) intervals, leading to a relatively small number of HCs that matched this inclusion criterion. Moreover, despite most MRIs from pwMS being conducted under observational studies (248/297 = 83.5%), the frequent scanning raises concerns about the generalizability of the findings in the present study to a broader MS population. Future, prospective replications of our results are needed to better understand how short-term susceptibility dynamics relate to disease progression.

## Conclusion

Our novel network-based analysis technique, IRONMAP, uncovered short-term, disease-related magnetic susceptibility abnormalities that were undetectable using a conventional per-region rate-of-change approach. IRONMAP has potential application for studying over-time brain iron abnormalities in a wide variety of neurological diseases, such as MS, Alzheimer's disease, and Parkinson's disease, over shorter timeframes than previously possible.

## Data and Code Availability

The data and code used to generate the final results is available upon reasonable request to the corresponding author.

## Author Contributions

## Funding

Research reported in this publication was supported by grants from the National Institutes of Health (R01NS114227 from the National Institute of Neurological Disorders and Stroke and UL1TR001412 from the National Center for Advancing Translational Sciences). The content is solely the responsibility of the authors and does not necessarily represent the official views of the National Institutes of Health.

## Declaration of Competing Interests

R.Z. has received personal compensation from Bristol Myers Squibb, EMD Serono, Sanofi, Mapi Pharma, Sana Biotechnologies and Filterlex for speaking and consultant fees. He received financial support for research activities from Bristol Myers Squibb, Mapi Pharma and Protembis and Filterlex.

M.D. received personal compensation from Bristol Myers Squibb, Novartis, EMD Serono and Keystone Heart, and financial support for research activities from Bristol Myers Squibb, Novartis, Mapi Pharma, Keystone Heart, Protembis, and V-WAVE Medical.

B.W.G. has participated in speakers bureaus for, served as a consultant for, and/or received research support from Biogen, EMD Serono, Novartis, Genentech, Celgene/Bristol Meyers Squibb, Sanofi & Genzyme, Janssen, Horizon, Bayer, and LabCorp. Dr. Weinstock-Guttman also serves on the editorial board for BMJ Neurology, Children, CNS Drugs, MS International, and Frontiers Epidemiology.

## Acknowledgements

The authors thank Ashley T., Alexander B., Zach W., and Daisy R. for their feedback on the manuscript.

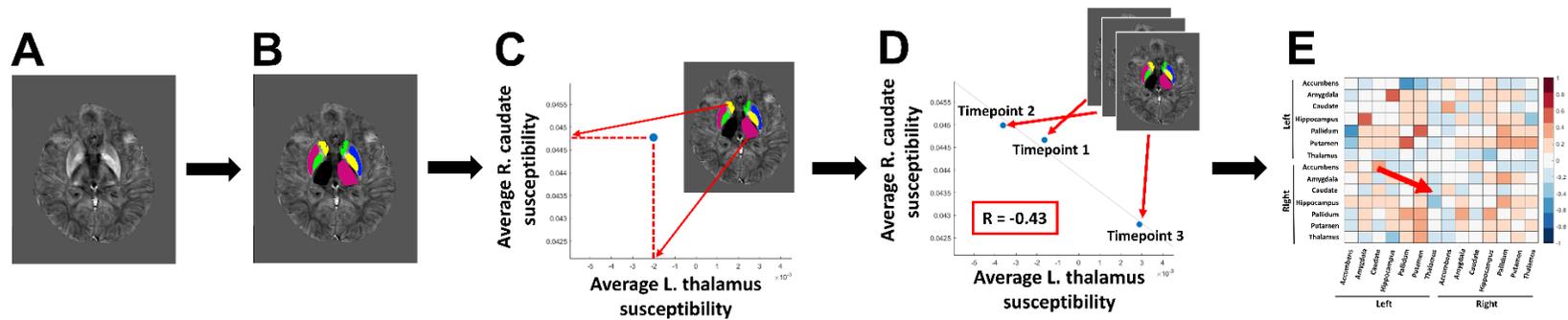

**Figure 1.** Procedure for the proposed network analysis. (A) Example of a susceptibility map. (B) Example of the corresponding deep gray matter segmentation. (C) Graph showing average susceptibility in the left thalamus compared to right caudate for a single subject at a single timepoint. (D) Pearson correlation between the left thalamus and right caudate for a single subject across three timepoints. (E) Group-average correlation coefficients organized into a correlation matrix. The red arrow in (E) points to the correlation coefficient generated in (D).

**Table 1. Demographic information for the people with MS and healthy controls.**

| Characteristic | pwMS | HCs | p-value |
|---|---|---|---|
| No. | 99 (85 pwRRMS, 14 pwSPMS) | 29 | |
| Age, yrs (mean±SD) | 43.7 ± 11.2 | 44.1 ± 15.6 | 0.764[a] |
| Sex, no. | | | 0.654[b] |
| Female | 66 | 21 | |
| Male | 33 | 8 | |
| dd, yrs (mean±SD) | 12.0 ± 8.8 | | |
| EDSS (median [IQR]) | 3.0 [1.5 – 4.5] | | |
| Number of gadolinium administrations over study timeframe (mean±SD) | 1.6 ± 0.7 | | |

**Legend:** *dd*- disease duration; EDSS- expanded disability status scale; pwMS- people with multiple sclerosis; HC- healthy control; pwRRMS- people with relapsing-remitting multiple sclerosis; pwSPMS- people with secondary progressive multiple sclerosis, SD – standard deviation.

[a]Two-tailed independent-samples T-test, [b]Chi-squared test.

**Table 2. Baseline and longitudinal susceptibility levels in DGM regions.** P-values < 0.05 are bolded.

|  |  | Baseline susceptibility (ppb) | | | Rate of susceptibility change (ppb/year) | | |
|---|---|---|---|---|---|---|---|
|  |  | pwMS | HCs | p-value* | pwMS | HCs | p-value* |
| Left | Caudate | 48.3 ± 14.6 | 41.3 ± 16.7 | **0.032** | 1.3 ± 29.7 | 4.1 ± 19.2 | 0.639 |
|  | Hippocampus | 6.2 ± 7.2 | 6.5 ± 8.1 | 0.870 | 1.4 ± 20.9 | -1.1 ± 11.0 | 0.544 |
|  | Pallidum | 115.5 ± 29.2 | 92.9 ± 28.1 | **< 0.001** | 8.9 ± 80.1 | -2.0 ± 25.2 | 0.473 |
|  | Putamen | 59.8 ± 18.7 | 54.4 ± 23.9 | 0.201 | 2.5 ± 47.3 | -2.3 ± 28.0 | 0.600 |
|  | Thalamus | 4.5 ± 9.7 | 8.0 ± 6.8 | 0.071 | -1.3 ± 18.0 | -1.8 ± 16.5 | 0.894 |
| Right | Caudate | 47.9 ± 15.8 | 39.2 ± 17.1 | **0.011** | 0.9 ± 33.6 | 1.8 ± 20.2 | 0.891 |
|  | Hippocampus | 6.0 ± 6.6 | 3.9 ± 7.0 | 0.149 | 0.9 ± 14.2 | -0.4 ± 8.7 | 0.631 |
|  | Pallidum | 107.6 ± 30.8 | 82.5 ± 29.8 | **< 0.001** | 10.2 ± 70.2 | -5.0 ± 28.9 | 0.260 |
|  | Putamen | 59.6 ± 19.5 | 54.1 ± 23.0 | 0.206 | 2.0 ± 52.4 | -5.9 ± 20.7 | 0.430 |
|  | Thalamus | 3.5 ± 9.0 | 5.7 ± 6.0 | 0.236 | -1.1 ± 18.2 | 1.3 ± 12.2 | 0.507 |

**Legend:** DGM – deep gray matter, HCs – healthy controls, ppb – parts per billion, pwMS – persons with multiple sclerosis.

*Two-tailed independent-samples T-tests.

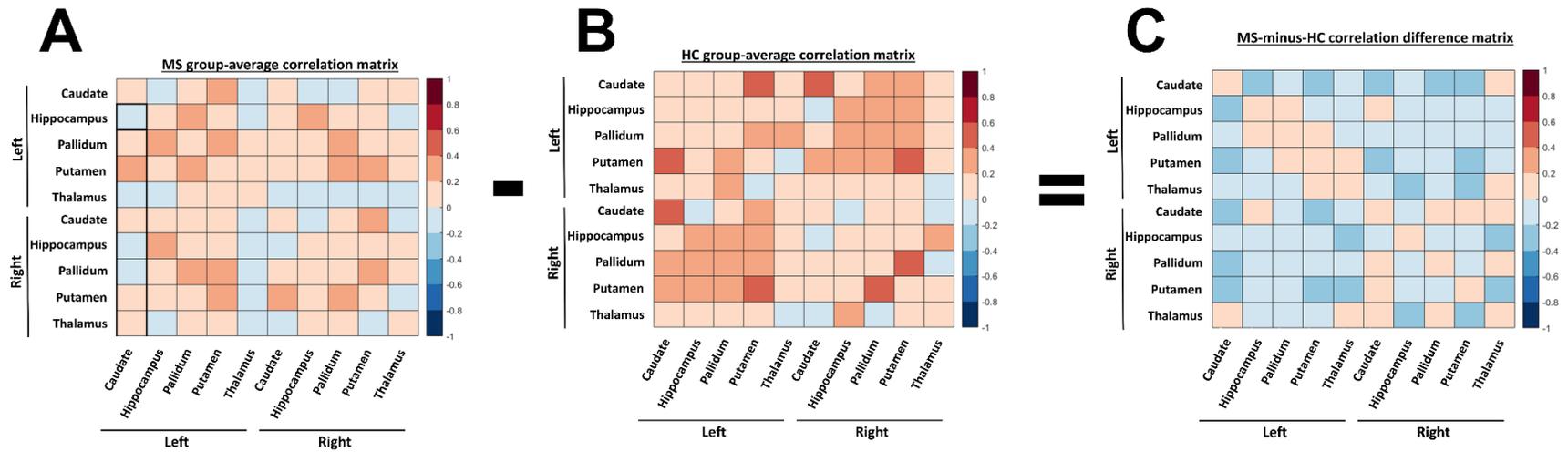

**Figure. 2.** Group average correlation matrices for (A) people with MS, (B) healthy controls, and (C) difference between the matrices of people with MS and HC correlation matrix.

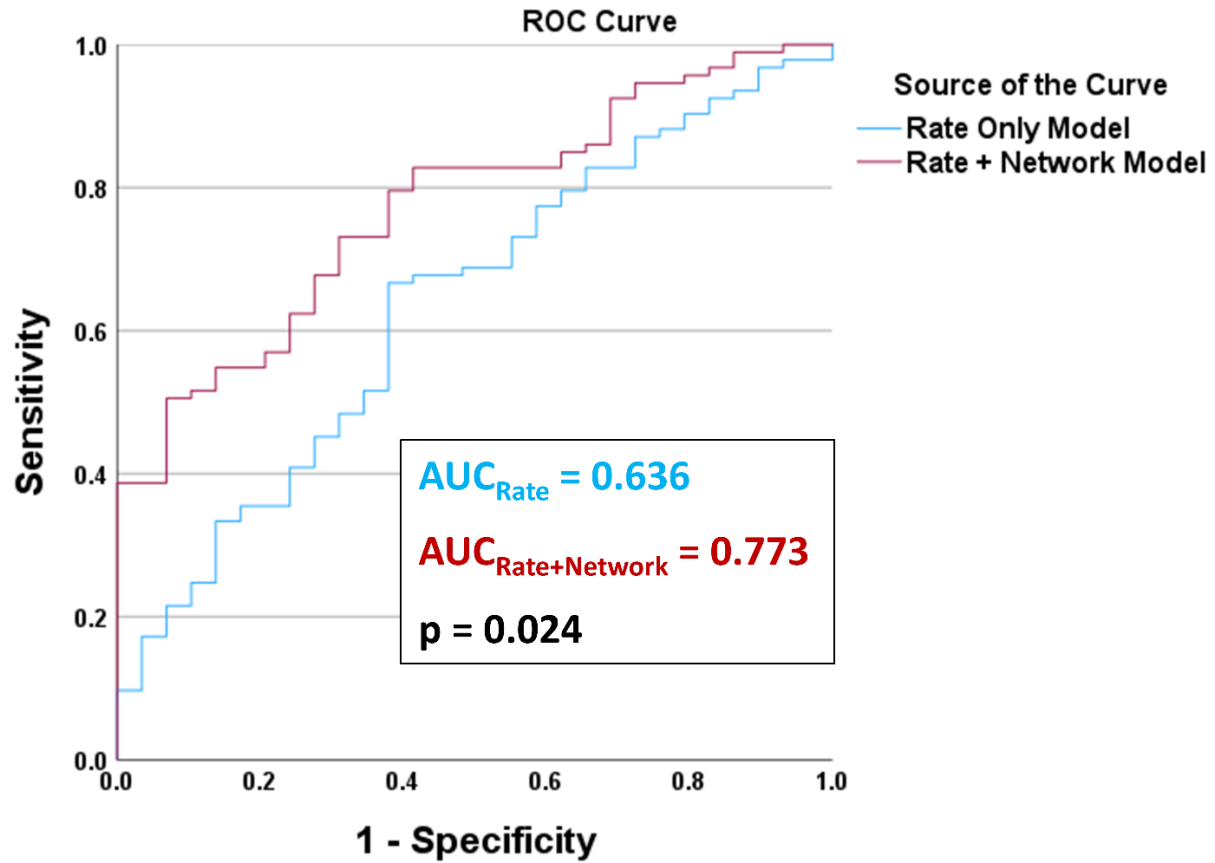

**Figure 3.** ROC curves for pwMS vs HC classification using a "Rate Only Model" and a "Rate + Network Model". Both models used binary logistic regression with subject group (pwMS or HC) as the outcome variable. The "Rate Only Model" model included the 10 per-region (i.e. five bilateral DGM structures) rates of susceptibility change as predictor variables. The "Rate + Network Model" added the 45 unique network region-pair correlations using forward selection (at $p < 0.05$), with the 10 per-region rates of susceptibility change included as forced entry predictor variables. The final "Rate + Network Model" included correlation coefficients between the left caudate and left pallidum, the left caudate and right caudate, and the right pallidum and right thalamus.

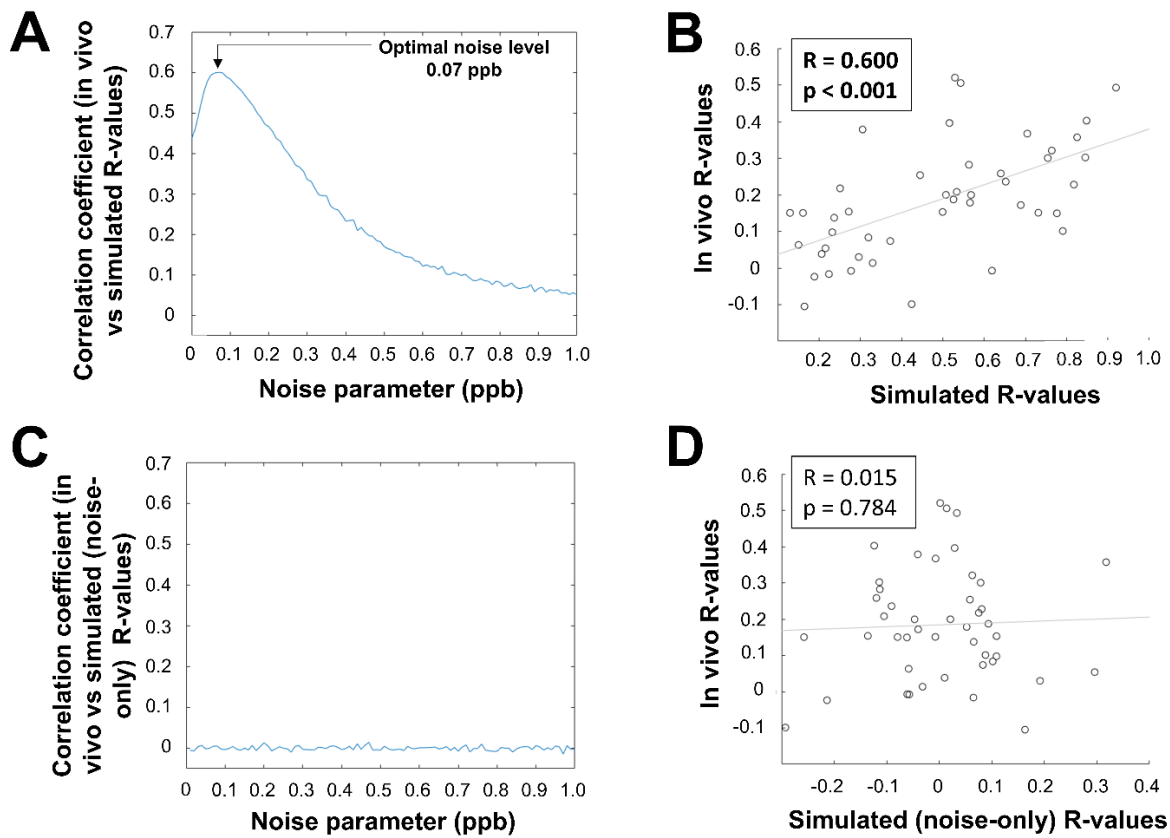

**Figure 4. Association between simulated correlation coefficients and in vivo data.** (A) Plot showing the relationship (correlation coefficients) between elements of the group-averaged HC in vivo network and corresponding elements of the simulated aging networks for different simulated noise levels, averaged over 1000 simulated iterations. (B) Scatter plot showing the group-averaged elements of the HC in vivo network (y-axis) and corresponding group-averaged elements of the simulated aging networks (x-axis) for the determined optimal noise level (0.07 ppb). (C) Plot showing the relationship (correlation coefficients) between elements of the group-averaged HC in vivo network and corresponding elements of the simulated noise-only networks for different simulated noise levels, averaged over 1000 simulated iterations. (D) Scatter plot showing the group-averaged elements of the HC in vivo network (y-axis) and corresponding group-averaged elements of the simulated noise-only networks (x-axis) for the determined optimal noise level (0.07 ppb).